\documentclass[prb,preprint,bibnotes,floats,aps,12pt]{revtex4}
\usepackage{times} 
\usepackage{graphicx}
\newcommand{\D}{{\rm d}}
\newcommand{\E}{{\rm e}}

\newcommand{\Ai}{{\rm Ai}}
\newcommand{\Bi}{{\rm Bi}}
\newcommand{\no}[1]{}
\newcommand{\ttfig}[1]{#1}
\begin{document}

\bibliographystyle{$LOCALHOME/Literatur/Bibdir/pof}
\title{Asymptotic theory for a moving droplet driven by a wettability
gradient}
\author{Len M. Pismen}
\email{pismen@techunix.technion.ac.il}
\affiliation{Department of Chemical Engineering and Minerva Center for Nonlinear Physics of
Complex Systems, Technion -- Israel Institute of
Technology, Haifa 32000, Israel}
\author{Uwe Thiele}
\email{thiele@pks.mpg.de}
\homepage{http://www.uwethiele.de}
\affiliation{Max-Planck-Institut f\"ur Physik komplexer Systeme,
N{\"o}thnitzer Str.\ 38, D-01187 Dresden, Germany}
\begin{abstract}
An asymptotic theory is developed for a moving drop driven by a wettability
gradient. We distinguish the mesoscale where an exact solution is known
for the properly simplified problem. This solution is matched at both -- the
advancing and the receding side -- to respective solutions of the problem on
the microscale. On the microscale the velocity of movement is used as the
small parameter of an asymptotic expansion. Matching gives the droplet shape,
velocity of movement as a function of the imposed wettability gradient
and droplet volume. 
\end{abstract}
%

%
%
\maketitle
%
%
%
\section{Introduction \label{intro}}

The description of the movement of a three-phase contact line is an
essentially unsolved hydrodynamical problem that
continues to attract much interest, for instance, when studying 
spreading drops, and liquid sheets or ridges moving down an inclined plate.
The understanding of 'simple' contact line movement is also paramount for a
deeper insight in related
problems as the dynamical wetting transition and
transversal instabilities of moving contact lines.

It is well known that the divergent shear stress at the contact line forbids a
solution in the framework of purely classical hydrodynamics, i.e.\ assuming a
no-slip boundary condition at the solid-liquid interface. Although this was
first pointed out by Huh and Scriven \cite{HuSc71} based on Moffatt's
\cite{Moff64} solution for flow in the edge which does not satisfy the normal
stress boundary condition on a free interface, and was never proven rigorously,
the divergence can be understood as a consequence of incompatibility of
multivaluedness of the velocity at the contact point in the classical
hydrodynamic formulation. 

The boundary condition has to be relaxed
to permit movement of the contact line. This can be done by introducing a very thin
precursor film on the 'dry' substrate \cite{SpHo96}, or by allowing for slip at the
solid-liquid interface everywhere \cite{HuSc71} or only near the contact line
\cite{Hock77,Gree78}, or introducing an
effective molecular interaction between the substrate and liquid into the
hydrodynamic model \cite{deGe85,TDS88}. For a discussion of the slip condition
see also the review by Dussan \cite{Duss79}.
Other approaches
include phase changes at the contact line \cite{Sepp96} or introduce
the vapour-liquid or fluid-solid interface,
or both, as separate phases with properties that differ from the bulk fluid
\cite{Shik97}. 

Most of the work on moving liquid sheets and ridges
prescribes a precursor film or slip at the substrate. Divergence problems at the
contact line are avoided, but at the expense of introducing ad hoc parameters into the
theory. These, namely the slip length or the precursor film thickness,
influence the profile of ridges and fronts and hence also the characteristics
of the transverse instability \cite{HoMi93,SpHo96,KaTr97,BeBr97}. 

The most realistic option is the explicit introduction of
molecular interactions into the hydrodynamic formalism. This is accomplished
by means of an additional pressure term, the disjoining pressure~\cite{DCM87}.
Depending on the particular problem treated, this disjoining pressure may
incorporate long-range van der Waals and/or various types of
short-range interaction terms~\cite{deGe85,TDS88,Isra92}.
Recently Pismen \cite{Pism01} derived a film thickness
equation with a disjoining pressure term 
by combining the long wave approximation for thin
films \cite{ODB97} with a nonlocal diffuse interface description
for the liquid-gas interface that incorporates  van der Waals interactions.

These interactions are essential for the process of dewetting, and studies
of dewetting of a thin liquid film on a substrate are generally based on models
involving a disjoining pressure \cite{Mitl93,ShKh98,Oron00,BGW01,TVN01,Beck03}.
Only a few studies of instabilities of liquid fronts have adopted
a similar approach \cite{ESR00,BeNe01,ThKn03}, despite the fact that such an approach
{\it predicts} all the ad hoc parameters of the slip or precursor models (i.e., the
static and dynamic contact angle, drop velocity, and the drop and precursor film thickness)
connected with the wetting properties of the liquid in terms of the parameters
characterizing the disjoining pressure. 

Recently, Eggers presented asymptotic solutions for the profile of advancing \cite{Egge04b}
and receding \cite{Egge04} driven contact lines (see also Ref.\,\onlinecite{Egge05}). 
The respective solutions match inner solutions near the contact line where a
slip model is used and outer solutions based on an analytic solution in terms
of Airy functions discussed in Refs.\,\onlinecite{BeOr78,Ford92,DuWi97,Voin97}.
However,  the advancing and receding case are studied for a plate pushed into
and pulled out of a liquid bath, respectively. It is not possible to directly
couple the two asymptotic solutions to describe the motion of a driven moving droplet
or ridge.

In the present work we tackle the problem of an asymptotic description of a
gradient-driven 
moving droplet that encompasses both an advancing and a receding contact
line. This implies that the description of the two contact lines and the
respective matching procedures depend on each other. 
Thereby we explicitly introduce the molecular interactions into the 
hydrodynamic formalism by using a chemical potential or disjoining
pressure describing a situation of partial wetting. This corresponds
to a precursor film model where the precursor film thickness
is determined through the disjoining pressure. 

We distinguish among three regions:
\begin{itemize}
\item \emph{Microscopic} (molecular scale) region: the dominant balance is
between disjoining potential and surface tension. 
\item \emph{Mesoscopic} region: the dominant balance is between viscous dissipation and surface tension.
\item \emph{Macroscopic} region: the dominant balance is between surface
tension and external forces 
\end{itemize}
Examples for driving forces are gravity for droplets or fronts on inclined
plates, Marangoni forces occurring if temperature gradients along the substrate
exist or wettability gradients along the substrate. Both, gravity and
Marangoni forces act in the lubrication limit as bulk forces, i.e.\ the force
is fed into the system in a top-down manner. This implies that the macroscopic
region has to be included in the description.
However, the third mentioned way to drive the system is based on a force
resulting from a wettability gradient that is fed into the system in a
bottom-up manner, i.e.\ on the microscale. The simplest description of such a
system is undertaken here by matching solutions obtained in the mesoscopic and
microscopic region. If the droplets are small enough (smaller than the
capillary length) the macroscopic scale can be ignored.

There are different physical situations where a gradient in wettability occurs
that can be mapped onto the presently studied model. (i) A droplet can 'sit' on a
step in wettability \cite{Raph88} allowing for an intermittent range of stationary 
movement until the complete drop sits on the more wettable substrate. (ii)
A droplet can move along a smooth wettability gradient
\cite{Broc89,DaCh02,PBGG04,DSGC04}. 
(iii) In a situation
involving an adsorption reaction at the substrate underneath the droplet a
droplet can produce the wettability gradient that drives its movement
\cite{DoOn95,LKL02,TJB04,JBT05}. 
In this way it carries the gradient along with its movement.
The latter case is also related to droplet motion caused by a 
surface phase transitions \cite{YoPi05}.

In the following we study all these situations in a model that uses a chemical potential
with different constants at the advancing and the receding contact
line, respectively. For situations (ii) and (iii) 
this corresponds to the assumption that the wettability gradient is small as
compared to the size of the contact zone but sizable as compared to the
overall droplet size.

In the next section the basic equations for the lubrication description of 
moving droplets are introduced. The exact solution in the mesoscopic region 
and its asymptotics are described in Section~\ref{S4}. The
microscopic solution and its asymptotic matching
are discussed in Section~\ref{S2}. 
Finally, a comparison of asymptotic and numerical results is given
together with our conclusions in section~\ref{concl}.

%
\section{Basic Equations \label{S1}}
%
Our starting point is the thin film evolution equation in lubrication approximation
\begin{equation}
\partial_t\,h = -\nabla\cdot \left\{ k(h)\, 
\nabla\, \left[ \gamma \epsilon^2 \nabla^2 h - \widehat{\mu}_s(h) \right]  \right\} .
\label{film}
\end{equation}
\no{film}
Here $\gamma$ is the surface tension of the liquid; $\epsilon$ is a scale
ratio used as a small parameter of the lubrication expansion (which will
further be identified with the local equilibrium contact angle). We shall use the
simplest mobility function $k(h)=\eta^{-1} h^3/3$, obtained under assumption
of constant dynamic viscosity $\eta$ with no slip at the substrate. The
chemical potential $\widehat{\mu}_s$ accounts for wetting properties. Note, that it
corresponds to the negative of a disjoining pressure $\Pi$ as used, for
instance, in Ref.\,\onlinecite{Shar93}.
For specific computations, we shall use the form \cite{Pism01}  
\begin{equation}
\widehat{\mu}_s(h)\,=\, \frac{Q_s}{h^3}\,\left[1-
\left(\frac{h_m}{h}\right)^3\right],
\label{mu}
\end{equation}
\no{mu}
where $Q_s$ is a characteristic excess fluid--substrate interaction energy,
which is proportional to the Hamaker constant \cite{Isra92}. If $Q_s>0$, this
form corresponds to a negative long-range and positive short-range part of the
spreading coefficient, thereby combining a destabilizing long-range and a
stabilizing short-range van der Waals interaction. The contact angle is
finite, and bulk fluid coexists at $\Pi=0$, i.e.\ in a flat layer of
macroscopic thickness in the absence of external forces, with an ultrathin
precursor of thickness $h_m$.

The variables in Eq.~(\ref{film}) are still dimensional but scaled to conform
with the lubrication approximation. They are related to the physical variables
(marked by a hat) as follows: 
\begin{equation}
\widehat{h}  = h, \qquad \widehat{x}  = x/\epsilon, \qquad  \widehat{t}  = t/\epsilon^2. 
\label{scale1}
\end{equation}
\no{scale1}
In consequence the scaled contact angle $\theta$ is related to the physical
one by $\theta=\widehat{\theta}/\epsilon$; the scaled droplet volume 
$V=\int h\,dx$ is related to the physical one
$\widehat{V}=\int\widehat{h}\,d\widehat{x}$ by $V=\epsilon\widehat{V}$.
Without any gradient parallel to the substrate, 
this model describes droplets with a finite equilibrium contact angle sitting
on an ultrathin precursor film.

However, here we are interested in moving droplets driven by wettability
gradients along the substrate. 
In the chemical potential chosen here [Eq.\,(\ref{mu})] a wettability increase can
be modelled by a decrease of $Q_s$ or by an increase of $h_m$. We chose here 
the former possibility. Note, however, that in a real physical system both
parameters are affected. The analysis then involves more algebra but is also
straightforward. 

We shall consider stationary motion of a 2D droplet with the velocity
$U$. Replacing in Eq.~(\ref{film}) $\partial_t h$ by $ - U h_{\overline x} /
\epsilon$ and integrating once yields, after dropping the bars,
\begin{eqnarray}
&& \frac{\mathrm{\delta}( h-h_m)}{h^3} =  
\frac{d}{dx} \left[  h''(x) - \mu_s(h) \right]\label{stat}\\
\qquad\text{with}
\qquad && \delta = \frac{3\mathrm{Ca}}{ \epsilon^3},
\qquad \mathrm{Ca} = \frac{\eta U}{\gamma},
\qquad \mu_s(h) = \frac{\widehat{\mu}_s(h)}{\gamma \epsilon^2}
\nonumber
\end{eqnarray}
\no{stat}
where $\delta$ is the appropriately rescaled capillary number Ca. 
For $\delta \ll 1$, this equation is
solved separately in the microscopic and mesoscopic regions, and solutions are
matched considering a respective subdominant term as a perturbation. Since
Eq.~(\ref{stat}) does not contain the coordinate explicitly, the order can be
further reduced (for a monotonic section) by replacing the variable $y(h) =
[h'(x)]^2$:
\begin{equation}
\pm  \frac{\delta( h-h_m)}{\sqrt{y}\, h^3}  = \frac{1}{2} y''(h)  -  
\mu_s'(h),  \qquad y(h_m)=0.
\label{staty}
\end{equation}
\no{staty}
In the next sections. solutions are determined in the mesoscopic and microscopic
region, respectively.
For comparison, the stationary moving droplets described by Eq.~(\ref{stat})
will also be computed numerically using  
continuation techniques \cite{DKK91,DKK91b} employing the
software AUTO97 \cite{AUTO97}. 

%
\section{Exact Mesoscopic Solution \label{S4}}
\subsection{General solution \label{S41}} 

At large distances ($h \gg 1$) a simplified ``mesoscopic'' equation can be
obtained by discarding the disjoining potential term in Eq.~(\ref{stat}) and
neglecting also $h_m \ll h$ in the viscous term:
\begin{equation}
\delta h^{-2} =  \partial_{xxx} h. 
\label{outout}
\end{equation}
\no{outout}
Rescaling the height $h= \delta^{1/3} \zeta$ reduces Eq.~(\ref{outout}) to
a parameterless form
\begin{equation}
\zeta^{-2} =  \partial_{xxx} \zeta. 
\label{out2}
\end{equation}
\no{out2}
This equation is invariant to \emph{simultaneous} rescaling of $\zeta$ and $x$. 
We chose $\delta>0$, however, results for $\delta<0$ can be obtained by the
transformation $x\rightarrow -x$.

Equation (\ref{out2}) 
has an exact solution expressed in a parametric form
through Airy functions \cite{DuWi97}: 
\begin{eqnarray}
\zeta(s) = \frac{K}{\pi^2 u^2(s)},  \qquad
x(s) &=& \frac{2^{1/3}K}{ u(s)}\,[\Ai(s)\Bi(s_0)-\Ai(s_0)\Bi(s)]
\nonumber\\[.3ex]
\mbox{with}\qquad
u(s) &=& \Ai(s)\Bi'(s_0)-\Ai'(s_0)\Bi(s) .
\label{airy}
\end{eqnarray}
\no{airy}

An indefinite factor $K$ appears here due to scale invariance of
Eq.~(\ref{out2}). It corresponds to the height of the droplet expressed
in units of $h_m$, i.e.\ it has to be large.

The parametric solution (\ref{airy}), generally, defines a discontinuous function
$\zeta(x)$, which is physically relevant only within certain intervals. 
For $s_0<s^\dagger$, where $s^\dagger 
\approx -1.01879$ is the largest zero of $\Ai'(s)$, physically irrelevant
solutions arise with $\zeta \to \infty$ for $x \to \pm \infty$ and a minimum
in between.  For $s^\dagger<s_0$ solutions exist with $h \to 0$ at $s \to
\infty$, which correspond to a sharp receding contact line at 
\begin{equation}
x^\star =
2^{1/3}K\,\Ai(s_0)/\Ai'(s_0) < 0.
\label{xstar}
\end{equation} \no{xstar}

\ttfig{
\begin{figure}[tbh]
\includegraphics[width=0.7\hsize]{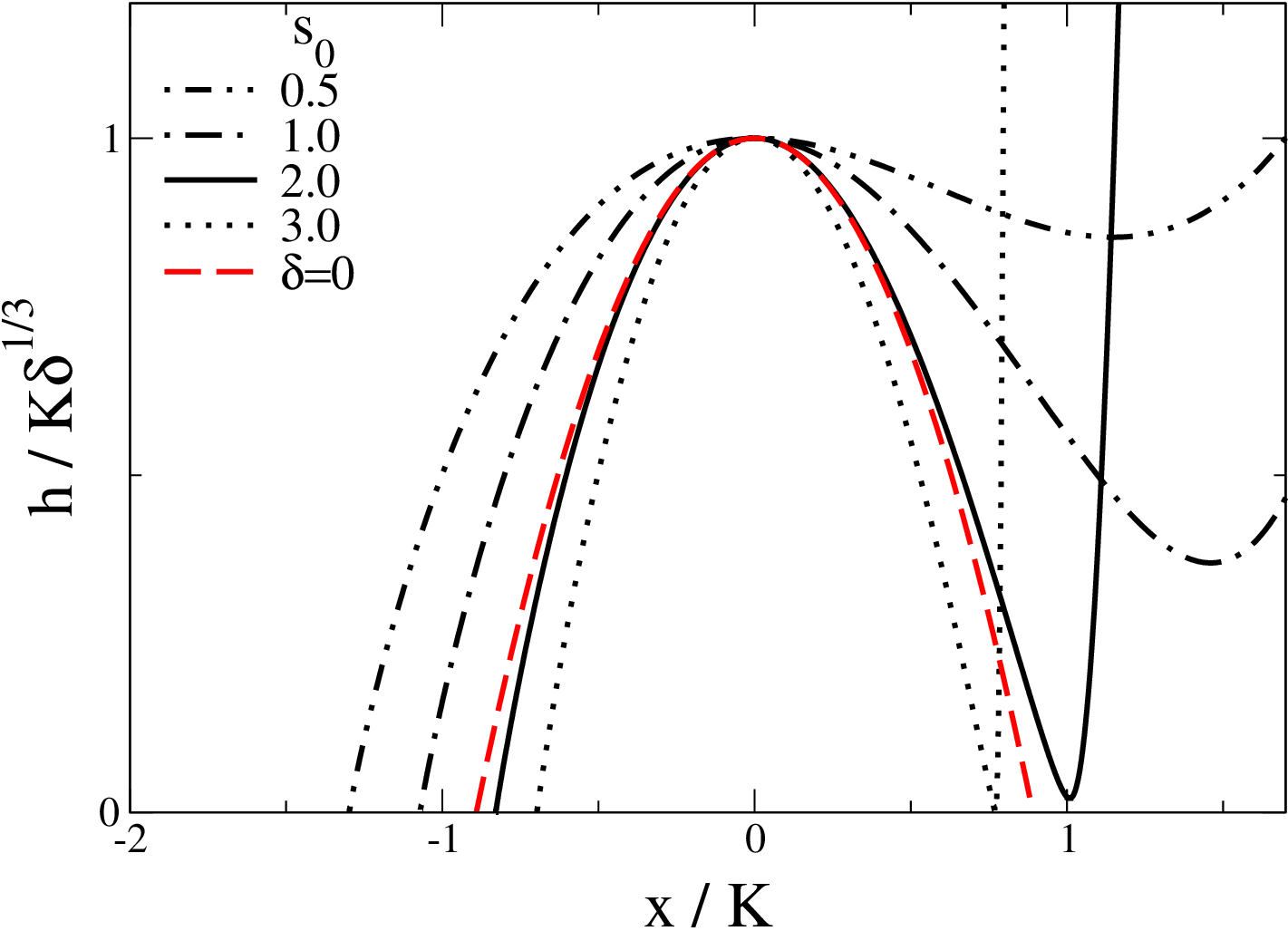}
\caption{Droplet profiles on the mesoscale given by
Eq.\,(\ref{airy}) for different $0\ge s_0$ as given in the legend. The
$y$-axis represents the height as $\zeta(x)/K$, i.e.\ corresponding to 
$h/K\delta^{1/3}$ in the scaling used on the microscale.
For comparison,
we also give the parabolic shape of a static droplet on a homogeneous
substrate. The droplet has the same volume 
$\int\zeta(x)dx$ as the one for $s_0=2$. %
\label{airyfig}\no{airyfig}
}
\end{figure}
}

For $s^\dagger<s_0<0$ the height $\zeta$ increases monotonically with
$x$.  These solutions are used as a model for a receding contact line in
Ref.\,\onlinecite{Egge04}. If, however, $s_0>0$, the profile $\zeta(x)$
has a maximum at $s=s_0$ corresponding to $x=0$; the solutions $\zeta(x)$ pass
through a minimum at $s_{min}<s_0$, i.e.\ $x_{min}>0$, before diverging as $\zeta \sim x^2, \; x
\to \infty$ at $s = s^\star$, where $s^\star(s_0) < 0$ is the largest zero of
$u(s)$.  As $s_0$ increases, the minimum comes very close to the $x$ axis 
and the curvature at the minimum becomes very large.
Examples of solutions for different $0\le s_0$ are shown in Fig\,\ref{airyfig}.

\subsection{Physically relevant interval \label{S42}} 

\ttfig{
\begin{figure}[tbh]
\includegraphics[width=0.7\hsize]{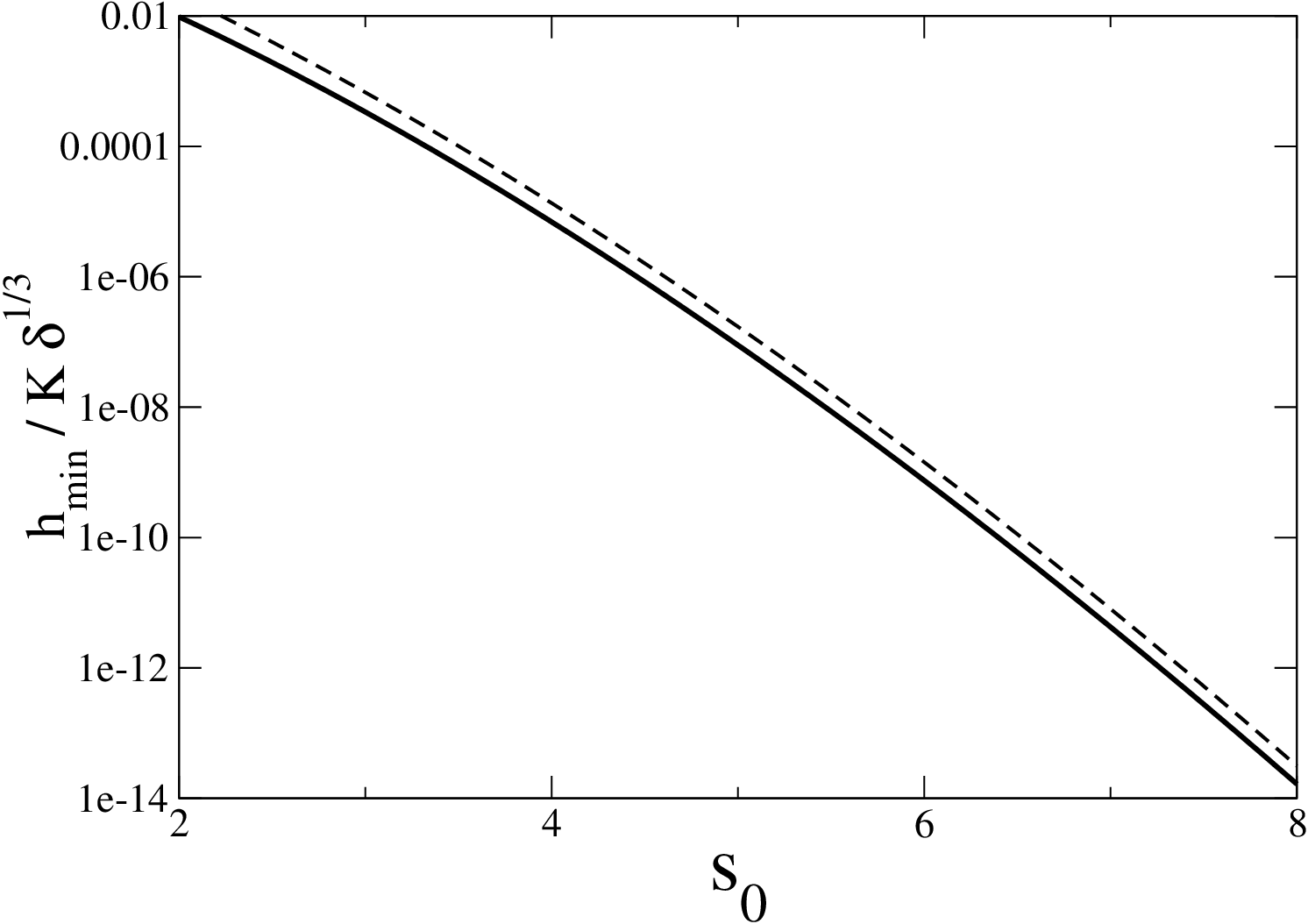}
\caption{The residual height at the minimum of the
droplet profiles as a function of $s_0$. The
ordinate represents the minimal height as $\zeta_{min}/K$, i.e.\ corresponding to 
$h_{min}/K\delta^{1/3}$ in the scaling used on the microscale. 
The solid line
gives the result using the approximative $s_{min}=-1.01879$ valid for
moderately large $s_0$ [Eq.\,(\ref{hmin})]. The dashed line
gives the result using the full Eq.\,(\ref{airy}).
\label{minheight}\no{minheight}
}
\end{figure}
}

We focus here on the case of moderately large $s_0$ taking the profile between 
$\zeta=0$ for $x=x^\star$ and the minimum of $\zeta(x)$ at $x=x_{min}$
as the outer solution for a moving droplet driven by a force fed in on the
microscale.  
The two parameters $s_0$ and $K$, as well as the droplet velocity that is
absorbed into the scaling, should be
obtained by matching the two inner (microscopic) solutions at
advancing and receding sides, as well as fixing the droplet volume.

For moderately large $s_0$ one finds that the location of the minimum  closely 
approaches $s_{min}= -1.01879$, which is the largest zero of  
$\Ai'(s)$. This follows from asymptotic relations applicable at moderately large $s_0$ 
that will be further discussed in Section~\ref{S5}. 
The resulting residual profile height at the minimum 
is plotted in Fig.~\ref{minheight}.  For a physical precursor film thickness
$h_m$ of the order of 1\,nm for a millimetric drop $K=10^6$ and one needs $s_0\approx4.5$.
For a droplet of one micron height $K=10^3$ and $s_0\approx2.5$.

Using the asymptotics of Airy functions at $s \to \infty$,
\begin{equation}
\Ai(s) \asymp \E^{-\frac{2}{3}s^{3/2}} s^{-1/4}\left[ \frac{1}{2 \sqrt{\pi}} \, + O\left(s^{-2}\right) \right], 
\qquad
\Bi(s) \asymp  \E^{\frac{2}{3}s^{3/2}} s^{-1/4}\left[ \frac{1}{ \sqrt{\pi}} \, + O\left(s^{-2}\right) \right] ,
\label{aba}
\end{equation}
\no{aba}
the solution can be expanded near the zero of $\zeta$ as
\begin{eqnarray}
x(s) &\asymp&  \frac{ 2^{1/3} K}{\Ai'(s_0)}  \left\{ \Ai(s_0) 
+ \frac{ \E^{-\frac{4}{3}s^{3/2}}}{2\pi \Ai'(s_0)} \left[1+ O\left( s^{-3/2} \right)  \right]   \right\}
+ O\left(\E^{-\frac{8}{3}s^{3/2}} \right), 
\label{xaia} \\
\zeta(s) &\asymp&  \frac{K \sqrt{s}\E^{-\frac{4}{3}s^{3/2}}  }{\pi\left[ \Ai'(s_0) \right]^2 }  
\left[ 1 + O\left( s^{-3/2} \right)  \right] + O\left(\E^{-\frac{8}{3}s^{3/2}} \right),
\label{hia}  
\end{eqnarray}
\no{xaia,hia}
Explicit asymptotics is obtained by solving Eq.~(\ref{xaia}) with respect to $s$:
\begin{equation}
s  \asymp \left( -\frac{3}{4}\ln \frac{x- x^\star}{L} \right) ^{2/3} 
    \left[ 1  +O \left(\ln^{-2}  \frac{x- x^\star}{L}  \right)  \right] .
 \label{hps}
\end{equation}
\no{hps}
where $L^{-1} =2^{2/3}\pi \Ai'(s_0)^2 /K  $. 
This yields, up to corrections of higher order in $\ln[(x- x^\star)/L]$, 
\begin{equation}
\zeta(x) \asymp  \left(x- x^\star  \right) \left(-3\ln \frac{x- x^\star}{L}  \right)^{1/3} 
, \quad
\zeta'(x) \asymp  \left(-3\ln \frac{x- x^\star}{L}  \right)^{1/3} .
 \label{hph}
\end{equation}
\no{hph}
The length $L$ is very large for moderately large $s_0$.

\subsection{Limit of weak driving \label{S5}} 

Although Eq.~(\ref{out2}) does not contain the rescaled capillary number
$\delta$, we expect  the applicable outer solution to
become symmetric, approaching a parabolic profile, 
in the limit  $\delta \to 0$, which corresponds to a vanishing wettability 
gradient. As illustrated in 
Fig.\,\ref{airyfig} and confirmed by the following asymptotic analysis,
the outer solution becomes almost symmetric at large values of $s_0$,
which, as we shall further see, correspond to small values of  $\delta$.

The limit $s_0 \to \infty$ can be obtained with the help of
the asymptotics (\ref{aba}) of Airy functions, which is practically applicable already at moderately large 
values $s_0 >2$. The resulting asymptotic profile height at the minimum on the advancing edge is
\begin{equation}
\zeta_{min}  \asymp K [ \pi  \Ai(s_{min}) \Bi'(s_0)]^{-2} 
 \asymp 1.10937 K s_0^{-1/2}\, \E^{-\frac{4}{3}s_0^{3/2}} \left[ 1 + O(s_0^{-3/2})\right] ,
\label{hmin}
\end{equation}
\no{hmin}
The minimum is located, up to an exponentially small correction proportional to 
$\E^{-\frac{4}{3}s_0^{3/2}}$, at the largest zero of $\Ai'(s)$, i.e. $s_{min}=
-1.01879$ (see Fig.~\ref{minheight}). 

The asymptotic expression for the second derivative $c=\zeta''(x)$ at the minimum is
\begin{eqnarray}
c(s_{min} )& \asymp& 2^{1/3} K^{-1}  \pi^2
 \left[ \Ai'(s_{min})^2 - s_{min} \Ai(s_{min})^2 \right] \Bi'(s_0)^{2} 
\nonumber \\
& \asymp& \widehat c K^{-1} s_0^{1/2} \, \E^{\frac{4}{3}s_0^{3/2}} \left[ 1 + O(s_0^{-3/2})\right] .,
\label{cmin}
\end{eqnarray}
\no{cmin}
with $\widehat c \approx 1.15697$.

The corresponding asymptotic value of the coordinate $x$ is
\begin{equation}
x_{min}  \asymp 2^{1/3} K  s_0^{-1/2}  \left[ 1 +  \mbox{$\frac{1}{4}$} s_0^{-3/2} + O(s_0^{-3})\right]  ,
\label{xmin}
\end{equation}
\no{xmin}
In the leading order, this coincides by the absolute value with the asymptotics of $x^\star$ 
given by Eq.~(\ref{xstar}):
\begin{equation}
x^\star  \asymp - 2^{1/3} K  s_0^{-1/2}  \left[ 1 -  \mbox{$\frac{1}{4}$} s_0^{-3/2} + O(s_0^{-3})\right] .
\label{xmin2}
\end{equation}
\no{xmin2}
This points out to the symmetry that should be attained in the limit of zero velocity. 
The full profile away from the location of the minimum should be computed 
by assuming \emph{both} $s_0$ and $s$ to be large. This yields, in the leading order, 
\begin{equation}
\zeta  \asymp  K (s/s_0)^{1/2} \: \mathrm{sech}^2 \left[  \mbox{$\frac{2}{3}$} (s_0^{-3/2} - s^{-3/2})\right] ,
\quad
x  \asymp 2^{1/3} K  s_0^{-1/2}  \tanh \left[  \mbox{$\frac{2}{3}$} (s_0^{-3/2} - s^{-3/2})\right] .
\label{xy0}
\end{equation}
\no{xy0}
At the receding edge, the asymptotics of these expressions at $s \gg s_0$ coincides with 
the asymptotics of  Eqs.~(\ref{xaia}), (\ref{hia}) at large $s_0$.

The height is of the same order of magnitude as the macroscopic length scale $K$
only when $s$ is close to $s_0$. Setting $s=s_0$, one can  see by combining the
above expressions that  
the function $\zeta(x)$ indeed approaches in this limit the parabolic profile $\zeta/K =
1-b^2 x^2$ with $b = 2^{-1/3}s_0^{1/2}/K$ everywhere except the immediate vicinity of both
contact lines; the corrections are of $O(s_0^{-2})$. Thus, the scaled droplet volume is computed as
\begin{equation}
V = 2K\delta^{1/3} \int_0^{1/b} (1-b^2 x^2 )\D x = \frac{4K\delta^{1/3}}{3b}
= \frac{4(2\delta)^{1/3}K^2}{3s_0^{1/2}}.
\label{vol0}
\end{equation}
\no{vol0}

\section{Microscopic Solution and Matching \label{S2}}
\subsection{Expansion in $\delta$ \label{S21}}

In the microscopic region, the thickness changes from $h = h_m$ to a
``mesoscopic'' value far exceeding $h_m$ but small compared to the drop size
and capillary length (that is here infinite). Solving Eq.~(\ref{stat}) with
$\delta=0$ defines the static contact angle in the limit $h \to \infty$, while for
$\delta\neq 0$ an \emph{apparent} dynamic contact angle is obtained in this
limit. The appropriate length scale in this region is $h_m$; the respective
dimensionless form of Eqs.~(\ref{stat}) and (\ref{staty}) is
\begin{equation}
 \delta \,\frac{h-1}{h^3} = \frac{d}{dx} \left[  h''(x) - \mu_s(h) \right],
\label{inner}
\end{equation}
\no{inner}
\begin{equation}
\pm \delta \,\frac{h-1}{\sqrt{y} h^3} = \frac{1}{2} y''(h)  -  \mu_s'(h), \label{innery}
\end{equation}
\no{innery}
 with 
\begin{equation}
  \mu_s(h) =\frac{\beta^2}{h^3}\,\left(1-\frac{1}{h^3} \right),  \qquad
  \beta= \frac{1}{\epsilon h_m}\, \sqrt{\frac{Q_s}{\gamma}} 
\label{scale3}
\end{equation}
\no{scale3}
To model different wettability at the advancing and receding contact line
one assumes different constants $\beta=\beta_{adv}$ and $\beta=\beta_{rec}$, respectively.
A higher wettability at the advancing side is assured by $\beta_{adv}<\beta_{rec}$.

For the receding meniscus, the positive sign should be chosen in
Eq.~(\ref{innery}), and the boundary conditions are $h = 1$, $h'(x) =0$ at $x
\to -\infty$, and $h''(x) \to 0$ at $x \to \infty$, 
 or $y(h)=0$ at $h=1$ and $y'(h)\to 0$ at $h\to\infty$.
The latter condition, suggested by Eggers \cite{Egge04},
should fit the curvature of the mesoscopic solution, which, according to
Eq.~(\ref{hph}), approaches $-\infty$ in the limit $x \to x^\star$.

For the advancing meniscus, the negative sign should
be chosen.  The boundary condition $h'(x) =0$ should be set at $x
\to \infty$, and the condition $h''(x) \to 0$ at $x \to -\infty$.

The solution of Eq.~(\ref{innery}) is sought for as an expansion in $\delta$:
$y(h)=y_0(h)+\delta y_1(h)+\ldots$. The zero-order equation,
\begin{equation}
 \frac{1}{2} y_0''(h)  -  \mu_s'(h)=0, 
\label{inner0}
\end{equation}
\no{inner0}
is readily integrated to obtain
\begin{equation}
 y_0(h) = \frac{3}{5} \,\beta^2\,\frac{(h-1)^2}{h^{5}}
 \left( \frac{2}{3} +\frac{4}{3}\,h +2h^2 +h^3 \right),
\label{yh0}  \end{equation} \no{yh0} 
The equilibrium contact angle $\theta_0$ is obtained from the zero-order
equation~(\ref{yh0}) in the limit $h \to\infty$: 
\begin{equation}
\theta_0 = h'(\infty)=  \sqrt{y_0(\infty)} = \sqrt{3/5}\, \beta.
\label{ythxi}  \end{equation}

The formal small parameter $\epsilon$ can now be identified with  the small 
physical equilibrium contact angle, say, $\widehat{\theta}_0^{rec}$ and expressed
through physical
parameters by requiring $\widehat{\theta}_0^{rec}/\epsilon=\theta_0^{rec} = 1$. This yields
\begin{equation}
\beta_{rec}=\sqrt{ \frac{5}{3} }, \qquad
\epsilon  = \widehat{\theta}_0^{rec} = \frac{1}{h_m} \,\left( \frac{3}{5} \,\frac{Q_s^{rec}}{\gamma} \right)^{1/2}  
 \propto \frac{d}{h_m} \, \left(\frac{Q_s^{rec}}{ Q_l} \right)^{1/2}.
\label{yth}  \end{equation}
The latter estimate follows from the estimate for surface tension $\gamma
\propto Q_l/d^2$, where $Q_l$ is a characteristic interaction energy of fluid
molecules and $d<h_m$ is the nominal molecular diameter. The contact angle is
indeed small when $Q_s^{rec}/ Q_l$ (the dimensionless Hamaker constant at the
advancing contact line) is small.
The numerical value of $\beta_{rec}$ is specific to the particular expression for
the disjoining potential (\ref{mu}), but the general procedure would be the
same for any potential of a similar shape. Note, that now only $\beta_{adv}<\sqrt{5/3}$
determines the driving wettability gradient.

Further derivation is carried out separately for receding and advancing menisci,
in view of different boundary conditions for the two cases.

\subsection{Receding meniscus \label{S22}}

The first-order equation derived from Eq.\,(\ref{innery}) is  
\begin{equation}
\frac{h-1}{\sqrt{y_0(h)}\,h^3} =  \frac{1}{2} y_1''(h) .
\label{inner1}
\end{equation}
\no{inner1}
Using here Eq.~(\ref{yh0}) and integrating from $h=1$ to $ \infty$ yields the
value of $y_1'(1)$ necessary to satisfy the asymptotic boundary condition of
vanishing curvature at $h \to \infty$ for the receding meniscus:
\begin{eqnarray}
 q_1   \equiv y_1'(1) & =& - 2 \sqrt{\frac{5}{3}}\frac{1}{\beta}\,\int_1^\infty 
     \left[ h \left( \frac{2}{3} +\frac{4}{3}\,h +2h^2 +h^3 \right)
\right]^{-1/2} \D h \nonumber\\
   &\approx& - 1.3383\,\sqrt{\frac{5}{3}}\frac{1}{\beta} = - 1.3383 . 
    \label{yhxi1}  
\end{eqnarray} \no{yhxi1}
The latter value corresponds to the scaling (\ref{yth}). 

A non-zero value of $y_1'(1)$ appears to change qualitatively the character of
decay to the equilibrium precursor thickness at very small deviations $h-1 \le
O(\delta)$.
At these distances, the expansion, in fact, breaks down, but the solution can
be readily found by linearizing Eq.~(\ref{inner}) near $h=1$. The linear
equation is solved by a combination of exponents $\E^{\lambda x}$ where
$\lambda$ is a positive root of $\lambda^3 - 3\beta^2 \lambda - \delta = 0$.
While for $\delta = 0$ the layer thickness decays at $x \to -\infty$ to unity as
$\E^{\sqrt{3}\beta x}$, for $\delta \neq 0$ an additional small root $\lambda
=  \delta \beta^{-2}/3 + O(\delta^2)$ appears.
This root is positive,
indicating a very slow decay to the equilibrium precursor thickness (and,
possibly, breakdown of quasistationary approximation) at $\delta \to 0$.

Since $y_1'(h) \sim h^{-1}$ at $h \to \infty$, $y_1(h)$ diverges logarithmically
in the outer limit. The asymptotic expression is obtained by integrating
Eq.~(\ref{inner1}) with the boundary condition (\ref{yhxi1}): 
\begin{equation}
 y_1 \asymp  - \,2 \ln   \frac{h}{a_1}, \qquad a_1 \approx 0.444 .
    \label{y1a}  \end{equation} \no{y1a}
The respective expansion for the slope $h'(x)$ useful for further matching to
an outer solution is
\begin{equation}
 h'(x) \asymp 1 - \delta \, \ln   \frac{ x}{a_1} + O(\delta^2) .
     \label{y1hx}  \end{equation} \no{y1hx}
The expansion can be routinely continued to higher orders with the help of a 
symbolic computation program.

To match the mesoscopic and microscopic solutions at the receding side, we compare
the outer limit of the receding microscopic solution 
($h\to\infty$) with the inner limit of the
mesoscopic solution ($\zeta\to0$). This translates to comparing
$[\zeta'(x)]^3$ given by Eq.~(\ref{hph}) with that given by Eq.~(\ref{y1hx}).
After rescaling Eq.~(\ref{hph}) and shifting the location of the contact line
to zero and rearranging Eq.~(\ref{y1hx}), this gives
\begin{equation}
[h'(x)]^3 =- 3\delta\,\ln \frac{x}{L}  =- 3\delta \ln \frac{\E^{-(1/(3\delta)}\,x}{a_1} .
 \label{hph1}
\end{equation}
\no{hph1}
The matching requirement yields the dependence of $L$ and, hence of $s_0$ on
$\delta$, expressed in an implicit form
\begin{equation}
L^{-1} = \frac{2^{2/3}\pi }{K} \, \Ai'(s_0)^2 
 \approx \frac{1}{2^{4/3}K\sqrt{s_0}} \,\exp \left[ -\frac{4}{3}s_0^{3/2} \right] = a_1
\exp \left[ - \frac{1}{3\delta}\right]  .
 \label{hph2}
\end{equation}
\no{hph2}
The approximate expression is valid for $s_0 \gg 1$ (practically, for
$s_0\gtrsim2$). This expression connects $s_0$ with the dimensionless velocity
$\delta$ as shown for different $K$ in Fig.\,\ref{delons0}.

\ttfig{
\begin{figure}[tbh]
\includegraphics[width=0.7\hsize]{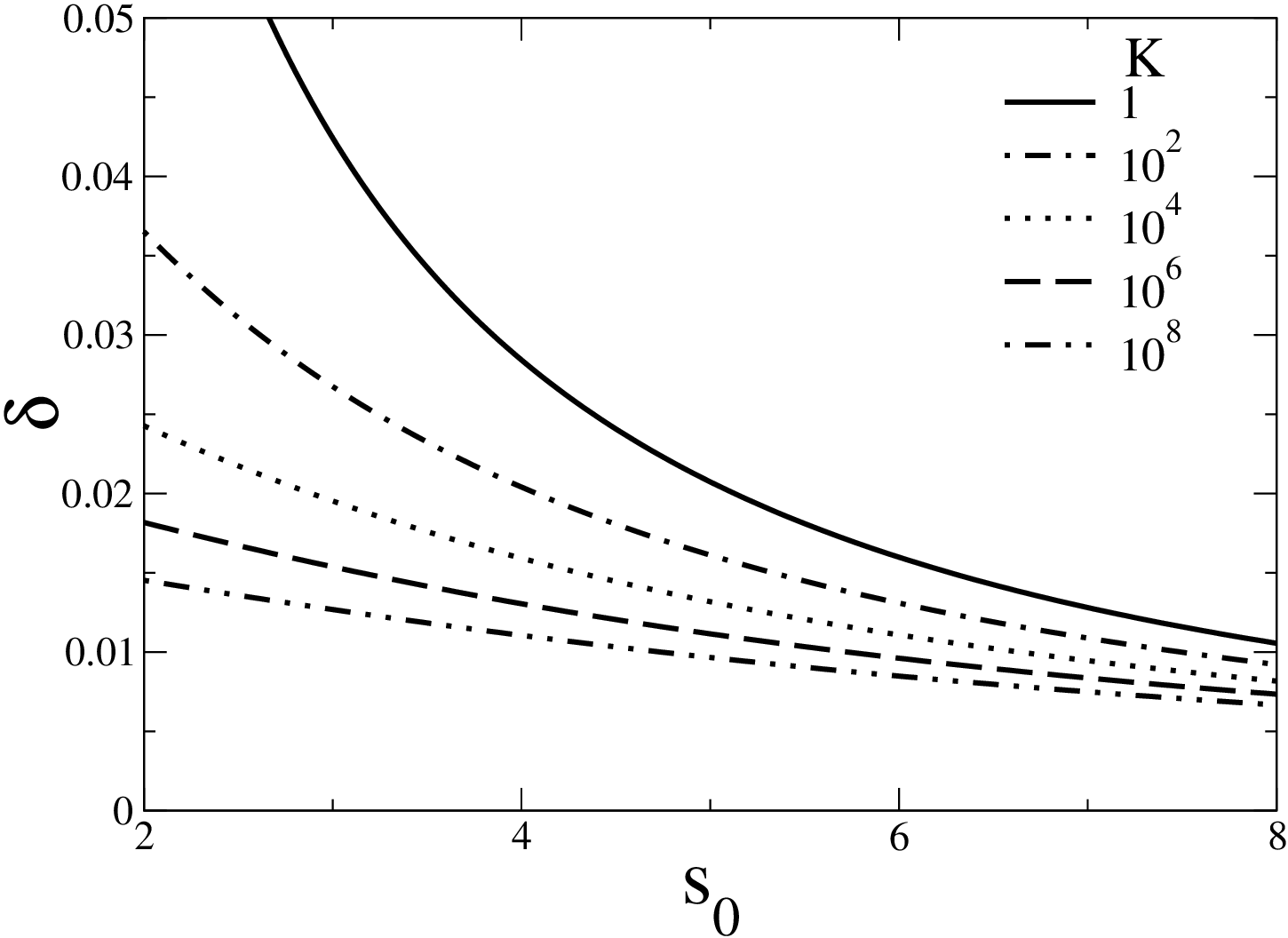}
\caption{The dependence of the rescaled capillary number $\delta$ 
on the parameter $s_0$ for different droplet sizes parametrized by $K$ 
(given in the legend) as given by Eq.\,(\ref{hph2}).
\label{delons0}\no{delons0}
}
\end{figure}
}
Combining this result with the dependence of $s_0$ on droplet size discussed
at Fig.\,\ref{minheight} gives an estimation of the velocity $\delta$.
For a physical precursor film thickness
$h_m$ of the order of 1\,nm for a millimetric drop $K=10^6$, $s_0\approx4.5$
and in consequence $\delta\approx0.01$
For a droplet of one micron height $\delta\approx0.025$.

\ttfig{
\begin{figure}[tbh]
\includegraphics[width=0.7\hsize]{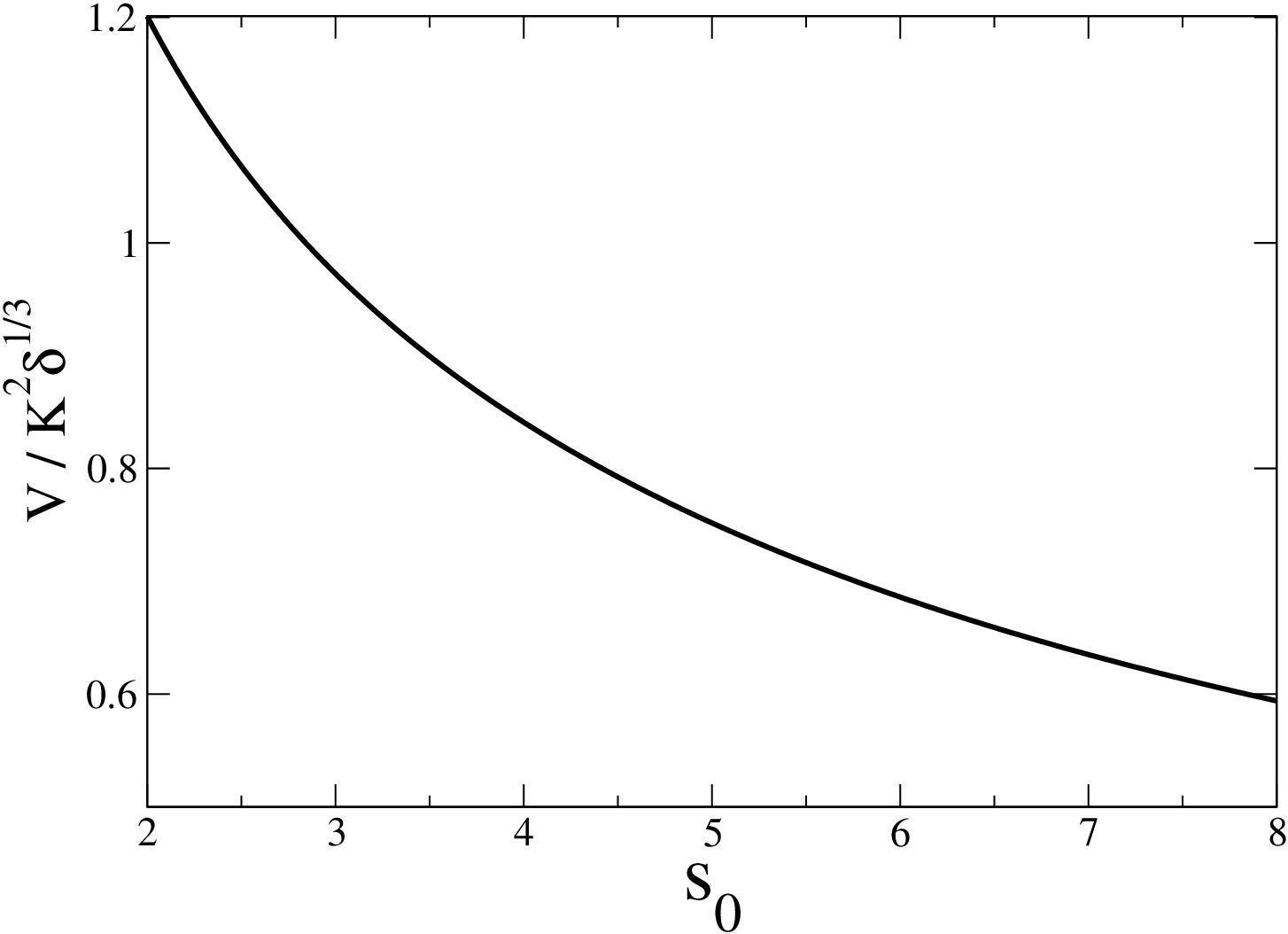}
\caption{Dependence of the droplet volume on the parameter $s_0$.
The $y$-axis represents the scaled volume 
$V/K^2\delta^{1/3}$ with $V$ given by Eq.\,(\ref{eqvol}).
$V/\theta_0$ represents a physical volume that may be used as a fixed
control parameter.
The numerical result using the minimum calculated with the full
Eq.\,(\ref{airy}) as integration boundary can not be distinguished from the solid line.
\label{vols0}\no{vols0}
}
\end{figure}
}
Note that this is still only an order-of-magnitude estimate, because
$K$ itself depends in a subtle way on the velocity. This can be seen in
Fig.\,\ref{vols0} where the dependency of the droplet volume
\begin{equation}
V\,=\,\delta^{1/3}K^2\,\int_{\infty}^{s_{min}} \zeta(s) x'(s) ds
\label{eqvol}
\end{equation}
\no{eqvol}
on $s_0$ is plotted.
To compare droplets of identical
volume for different driving forces, one has to determine $K$ using the
matching at the advancing edge.

\subsection{Advancing meniscus \label{S23}}

For an advancing contact line, the mesoscopic solution has no logarithmic asymptotics,
and for matching one can use the zero-order
microscopic solution, matching its limit at $h\to\infty$, $h''(x)\to 0$ to the mesoscopic 
solution at the inflection point $\zeta''(x) =0$.

This translates to comparing $\delta^{1/3}\partial_x\zeta(s_i)=\delta^{1/3}\zeta'(s_i)/x'(s_i)$ 
at the inflection point $s=s_i$ given by
\begin{equation}
s_i\,[\Ai'(s_0)\Bi(s_i)-\Ai(s_i)\Bi'(s_0)]^2\,=\,[\Ai'(s_0)\Bi'(s_i)-\Ai'(s_i)\Bi'(s_0)]^2
\label{infl}
\end{equation}
\no{infl}
to $h'(x\to\infty)=\theta_0^{adv}$ defined by Eq.\,(\ref{ythxi}) with $\beta$
replaced by $\beta_{adv}$.  
As result of the matching, one finds 
\begin{equation}
\delta\,=\,\left(\frac{\theta_0^{adv}}{\partial_x\zeta(s_i)}\right)^3,
\label{advmatch}
\end{equation}
\no{advmatch}
i.e.\ $\delta/(\theta_0^{adv})^3$ can be calculated as a function of the parameter $s_0$ as
presented in Fig.\,\ref{deltaadv}.

\ttfig{
\begin{figure}[tbh]
\includegraphics[width=0.7\hsize]{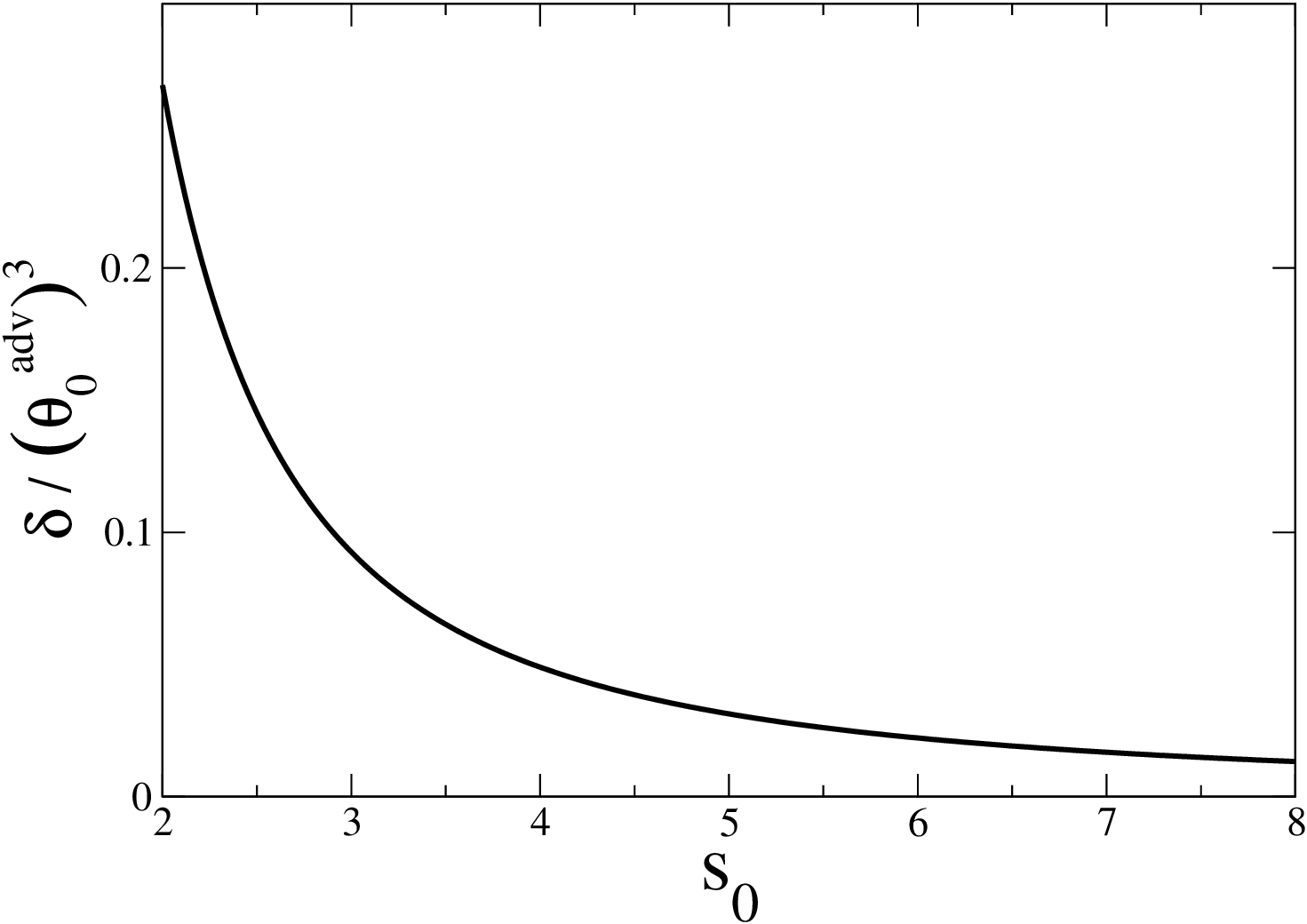}
\caption{Dependence of the scaled droplet velocity $\delta/(\theta_0^{adv})^3$ on the
parameter $s_0$, as obtained from the inflection point 
matching at the advancing contact zone [Eq.\,(\ref{advmatch})].
\label{deltaadv}\no{deltaadv}
}
\end{figure}
}

This procedure effectively cuts off the highly curved segment 
of the mesoscopic solution near the minimum. 
Take note that $\zeta_i$ is
still much larger than the $O(1)$ microscopic scale, and one
can expect corrections due to the disjoining potential to become
significant only well below this value.
However, as we will illustrate in the Conclusion, the first order matching is already 
sufficient to completely describe the droplets driven by a wettability 
gradient.

\section{Conclusion \label{concl}}

We have developed an asymptotic theory for a moving drop driven by a wettability
gradient. Wide separation between the meso- and microscale allows us to
use respective analytical and expanded solutions on the different scales.

Matching of the mesoscale and microscale solutions at the advancing and the
receding contact region allows to obtain the droplet shape and the
velocity of movement as functions of the imposed wettability gradient
and droplet volume. In this way, the two matching procedures together 
with a translation between the different
scalings gives a complete characterization of the droplet motion for
a given physical volume $V/\theta_0$ and the wettability gradient characterized
by the physical receding
$\hat{\theta}_0^{rec}=\theta_0\theta_0^{rec}=\theta_0$ 
and advancing $\hat{\theta}_0^{adv}=\theta_0\theta_0^{adv}<\theta_0$
equilibrium angle.

\ttfig{
\begin{figure}[tbh]
(a)\includegraphics[width=0.6\hsize]{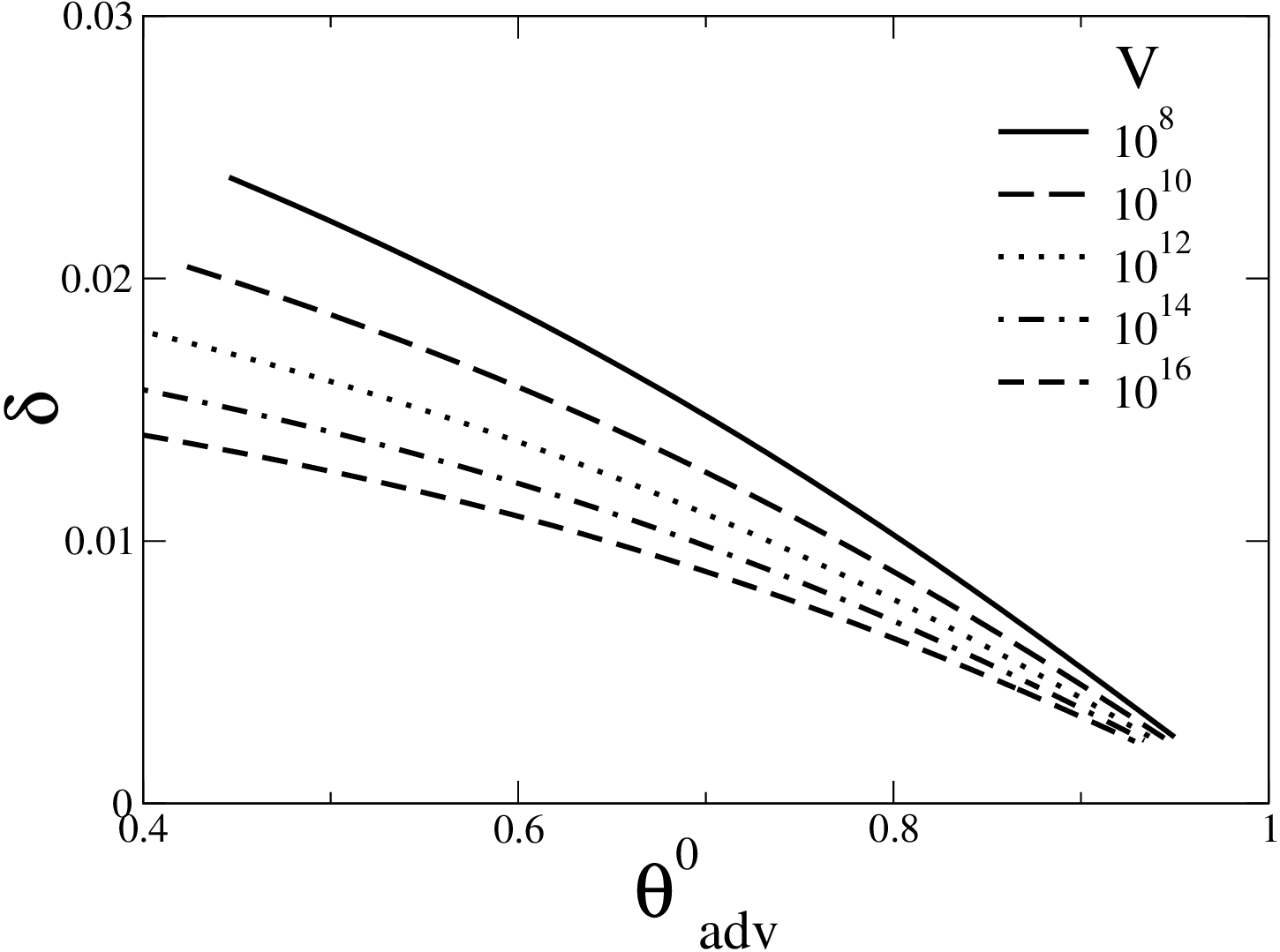}\\
(b)\includegraphics[width=0.6\hsize]{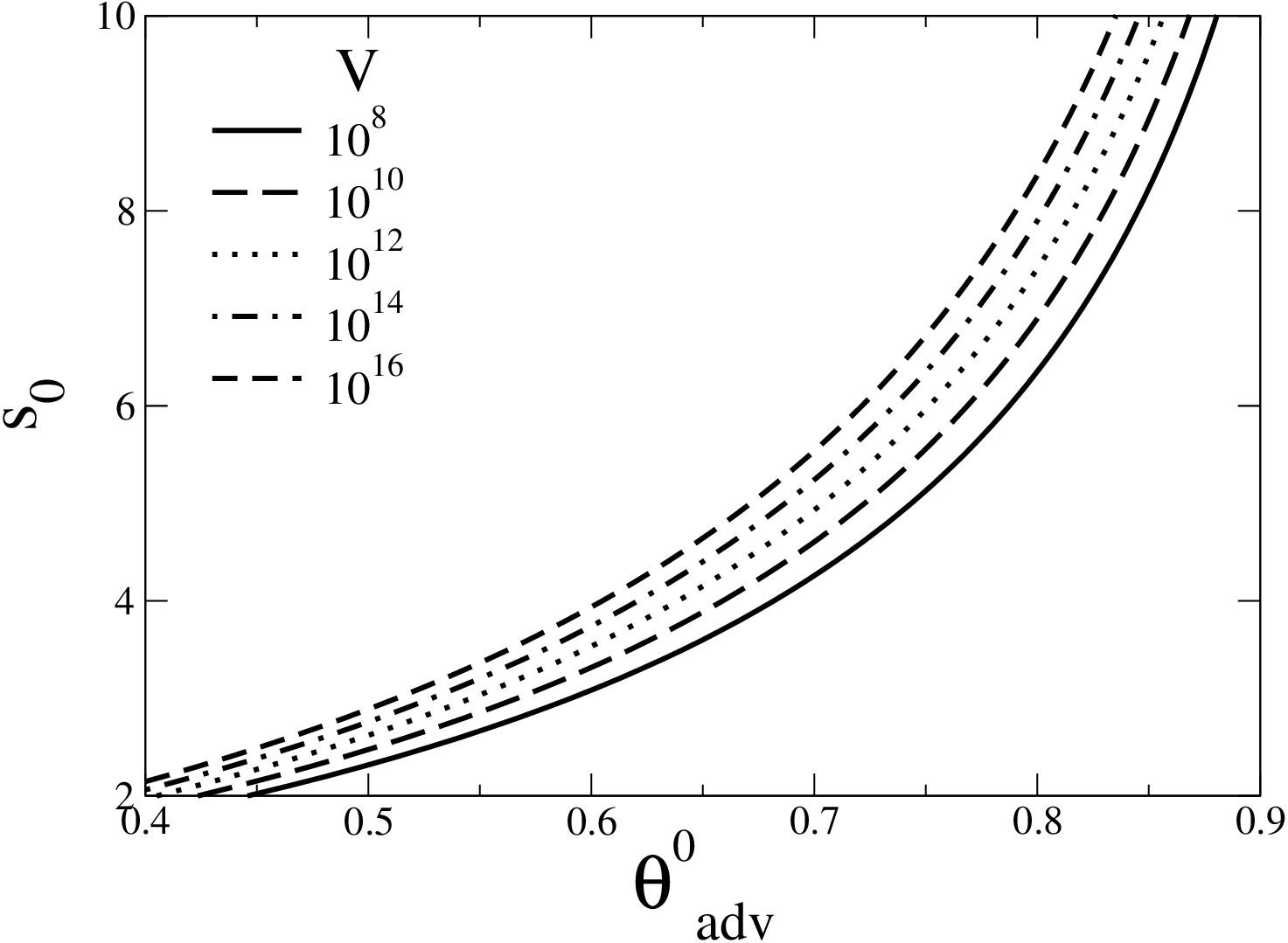}
\caption{Results of the numerical matching proceedure described in the main
text using at the advancing contact zone the inflection point matching. 
Shown are (a) the droplet velocity $\delta$, 
and (b) the parameter $s_0$
describing the mesoscopic shape in their dependency on the imposed
equilibrium advancing contact angle $\theta_{adv}<\theta_{rec}=1$ for
different given droplet volumes $V$ as given in the legend. In (a) results are
shown for $2\le s_0 \le 20$.
\label{numresults}\no{numresults}
}
\end{figure}
}

Fixing the overall long-wave scaling by fixing $\theta_0$, the 
three relations between
$\theta_0^{adv}$, $V$, $\delta$, $K$ and $s_0$ obtained in the course of the
present work [Eqs.\,(\ref{hph2}), (\ref{eqvol}) and (\ref{advmatch}) illustrated
in Figs.~\ref{delons0}, \ref{vols0} and
\ref{deltaadv}, respectively] allow us to determine the unknown $\delta$, $K$
and $s_0$ for each given pair of $\theta_0^{adv}$ and $V$.
In Fig.~\ref{numresults} results of the asymptotic matching are given 
for the velocity $\delta$, 
and the ``shape parameter'' $s_0$ in dependence of the advancing equilibrium 
contact angle $\theta_0^{adv}$ for a selection of volumes $V$. As expected, the
velocity goes towards zero as the driving wettability difference
$\theta_0^{rec}-\theta_0^{adv}$ vanishes, i.e.\
$\theta_0^{rec}\rightarrow 1$. The shape parameter $s_0$ diverges for
$\theta_0^{rec}\rightarrow 1$ as discussed in section \ref{S5}.
At a fixed driving $\theta_0^{adv}$, the droplet becomes more asymmetrical
($s_0$ decreases) and faster with decreasing volume. The velocity changes with volume
are more pronounced for larger driving (i.e.\ smaller $\theta_0^{adv}$).

Albeit the matching is based on an expansion in $\delta$, the numerical
calulations leading to Fig.\,\ref{numresults} are not practical for very
small $\delta$ (i.e.\ large $s_0$) because already for $s_0=20$ the
calculation involves small numbers of the order of $e^{-120}$ that are difficult to
handle. Under these conditions, the shape, however, remains almost static, and the integral
relations of Ref.~\onlinecite{PiPo04} can be used. 
The relations obtained by
multiplying Eq.~(\ref{stat}) by $h-h_m$ and integrating over the entire $x$
axis yield the expression for the dimensionless velocity in the form of a
ratio $\delta \,=\, F/I$
of the driving force $F$ to the dissipative integral
\begin{equation}
I \,=\,  2 \ln \frac{2a}{bh_m},
\end{equation}
where $a=(3/2\,V)^{1/2}$ is the radius of a static parabolic droplet with the profile 
$h = \frac{1}{2} a [1-(x/a)^2]$ and $b \approx 2.082$ is a constant. 
The driving force $F=F^{rec}-F^{adv}$ is expressed through the equilibrium
contact angles by separating the contributions of the two menisci $F^{rec},
\;F^{adv}$: 
\begin{equation}
F  = - \int_{-\infty}^{\infty} (h-h_m)\frac{d\mu_s}{dx} dx = \int_{-\infty}^{\infty} \mu_s \frac{dh}{dx} dx 
= F^{rec} - F^{adv}, 
\end{equation}
where, after replacing the integration variable and extending integration to
infinity 
in a thick middle part of the droplet where the disjoining potential is negligible,
\begin{equation}
F^{rec,adv}  =  \int_{h_m}^{\infty} \mu_s^{rec,adv} (h) dh =   \frac{\widehat \theta^2_{rec,adv}}{2}.
\end{equation}
This yields (with $ \theta_{rec}=1$) 
\begin{equation}
\delta = \frac {1- \theta_{adv}^2}{4 \ln (2a/bh_m)}.
\label{pp04delta}
\end{equation}
This result is compared to the asymptotic theory in Fig.\ref{fincomp}.

The presented asymptotic theory is based on (i) a separation into micro- and
mesoscale, i.e.\ it is not valid for $V$ too small ($s_0$ becomes too
small) and (ii) an expansion in $\delta$, i.e.\ it is not valid for $\delta$ 
too large. Assuming a precursor film of 1~nm, $V=10^6$ corresponds
roughly to droplets of 1\,$\mu$m height, implying that the asymptotics is
valid in the realm of microfluidics, but less so for nanofluidics.
However, because for 
nanodroplets the micro- and mesoscale are not well separated they can be
treated with numerical methods. It is convenient to 
calculate stationary moving droplets using continuation techniques 
\cite{DKK91,DKK91b} as shown, for instance, for nanodroplets moving under the
influence of a body force \cite{Thie01,Thie02} and chemically driven droplets 
\cite{TJB04,JBT05}. However, the numerical calculation becomes very tedious
for larger drops because of the separation of scales.

\ttfig{
\begin{figure}[tbh]
\includegraphics[width=0.7\hsize]{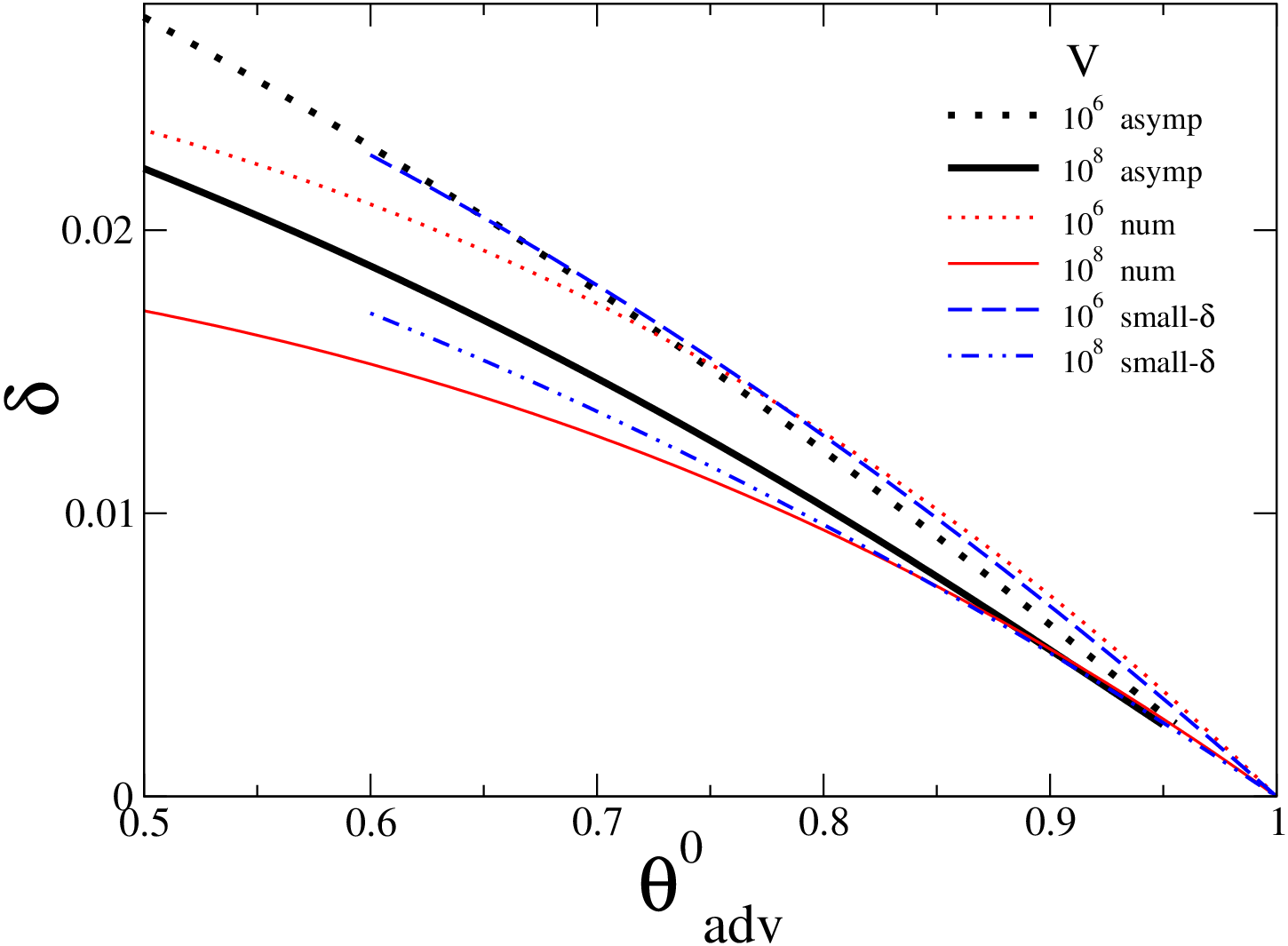}
\caption{Comparison of asymptotic results (thick lines) obtained from 
Eqs.\,(\ref{hph2}), (\ref{eqvol}) and (\ref{advmatch}),
numerical continuation results (corresponding thin lines) for
Eq.\,(\ref{inner}), and the small-$\delta$ results given by
Eq.\,(\ref{pp04delta}) (dashed lines).
The dependence of the droplet velocity $\delta$, 
on the imposed equilibrium contact angle at the advancing side 
$\theta_{adv}<\theta_{rec}=1$ is shown for
different given droplet volumes $V$, as specified in the legend. 
\label{fincomp}\no{fincomp}
}
\end{figure}
}

In Fig.\,\ref{fincomp} we present a comparison of asymptotic results (stretched
down to $V=10^6$) obtained from Eqs.\,(\ref{hph2}), (\ref{eqvol}) and
(\ref{advmatch}), small-$\delta$ results given by Eq.\,(\ref{pp04delta}),
and numerical continuation results (stretched up to  $V=10^8$) for Eq.\,(\ref{inner}).
For small driving $\theta_{adv}^0>0.8$ the overall agreement of the three
methods is reasonably good. For $V=10^6$ the maximal deviation is below
15\%, and for $V=10^8$ it
is about 5\%. As expected, for larger driving $\theta_{adv}^0<0.8$ the results
start to deviate, the numerical solutions of the full Eq.\,(\ref{inner}) give
a lower velocity than the asymptotics, and more so for smaller
$\theta_{adv}^0$. For larger droplets this deviation starts at larger  
$\theta_{adv}^0$ (smaller driving). 
There are various small factors that may contribute to
the deviations at small driving: (i) for $V=10^8$ 
the equilibrium contact angle still differs from the asymptotic value of one 
by about 0.3\%;
(ii) for moving droplets the precursor film thickness depends weakly on the dynamics 
\cite{Thie01,ThKn04} implying a droplet volume that is not exactly 
constant with changing velocity. For $V=10^8$ and 
$\theta_{adv}^0=0.5$ the precursor film 
thickness is about 1.0025, i.e.\ for the used domain size of $10^6$ the 
relative change in droplet volume is negligible ($\Delta V/V\approx10^{-5}$).

Surprisingly, the simple results obtained for small $\delta$ 
in Ref.~\onlinecite{PiPo04} as the ratio of 
the driving force and the dissipative integral [our Eq.\,(\ref{pp04delta})]
seem to fit the numerical data better than the asymptotic theory.
This results, apparently, from the cancelation of different approximations.
The assumed velocity-independent parabolic droplet shape underestimates, for instance,
the dissipation at the receding contact line and in the bulk, but overestimates the 
dissipation at the advancing contact line. The advantage of the asymptotic 
theory can be better appreciated comparing the profiles of the moving droplets
(Fig.\ref{compprof}).

\ttfig{
\begin{figure}[tbh]
(a)\includegraphics[width=0.6\hsize]{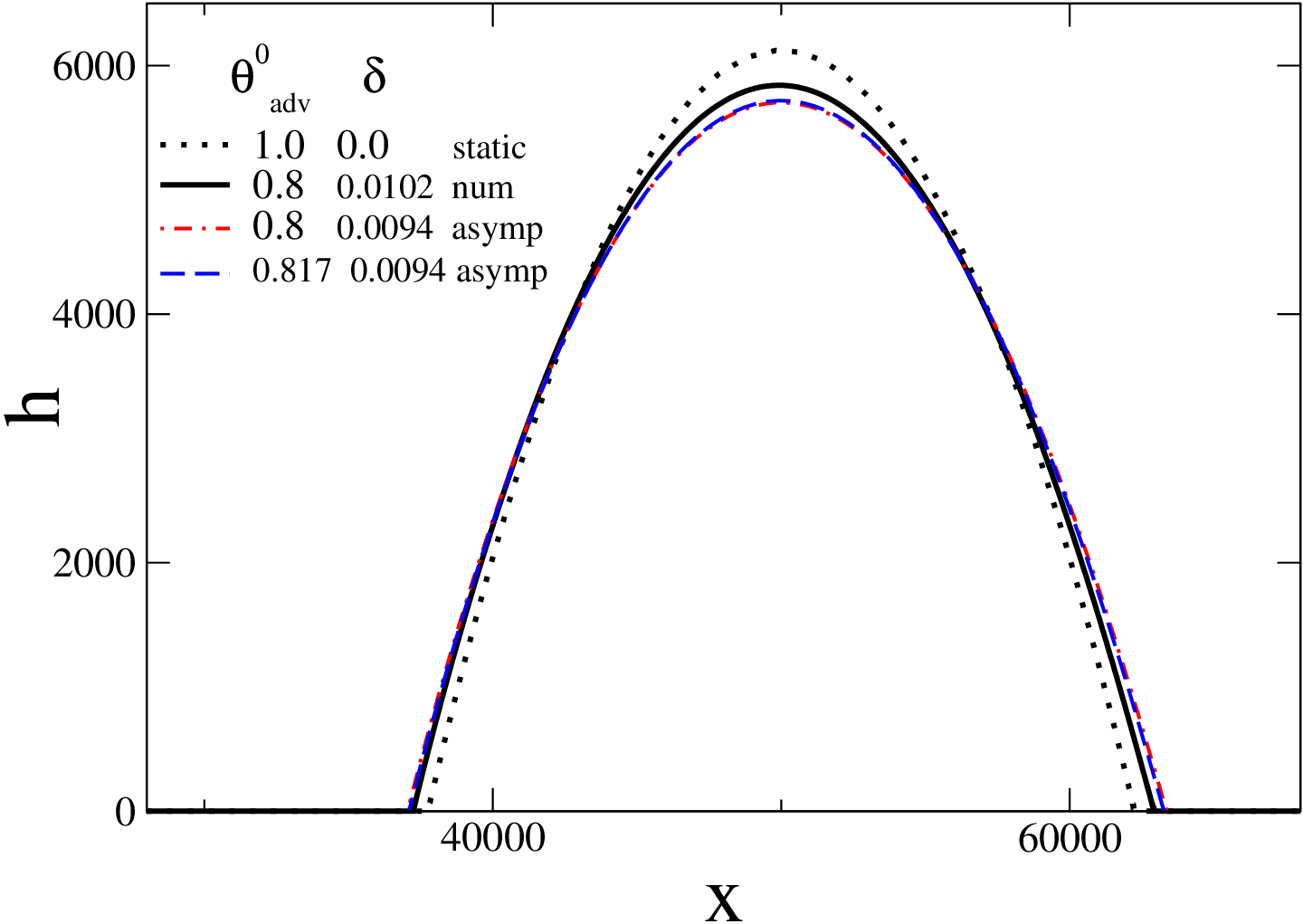}
(b)\includegraphics[width=0.6\hsize]{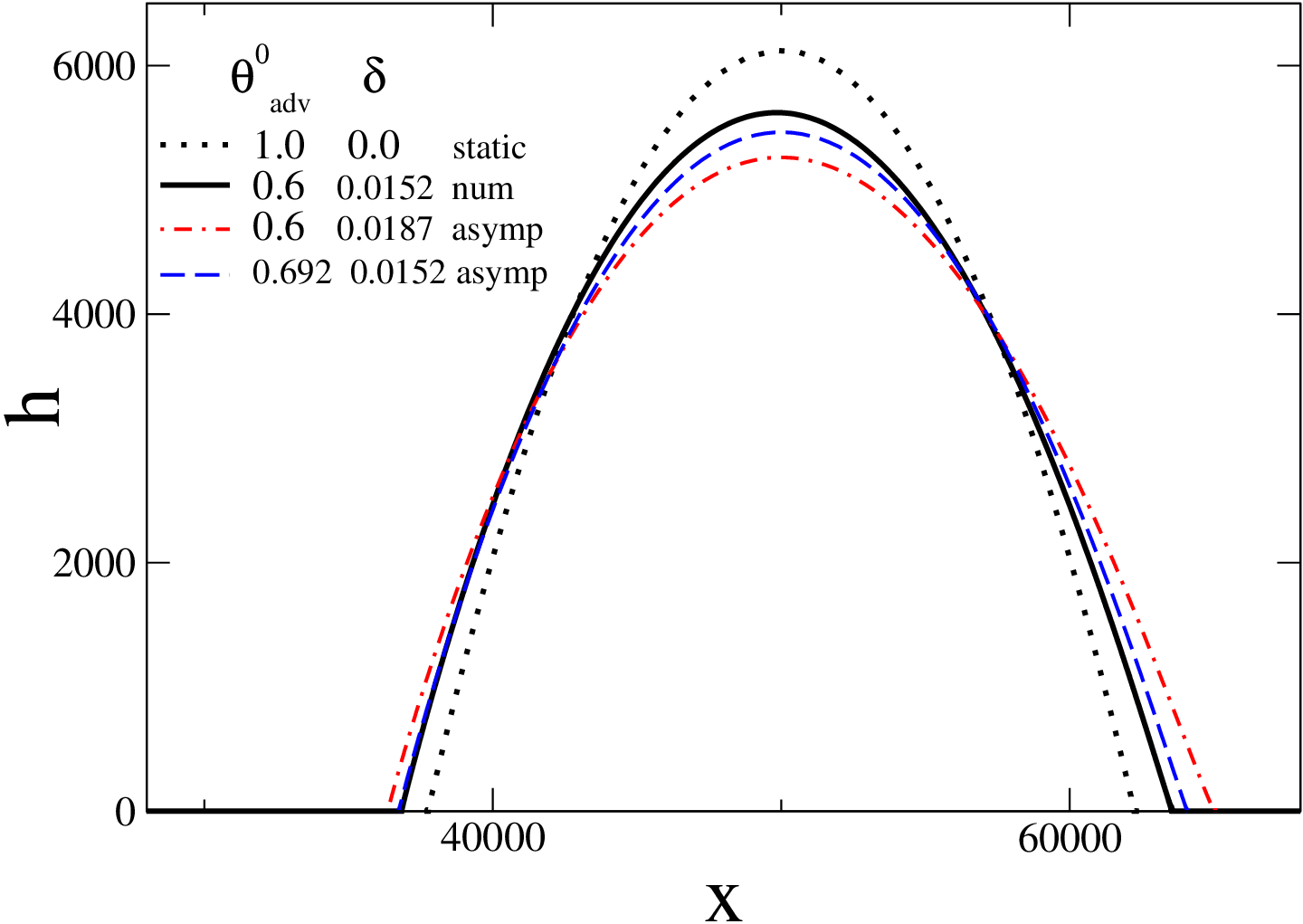}\\
\caption{Profiles of moving droplets for $V=10^8$ as obtained by numerical 
continuation of stationary solutions [Eq.\,(\ref{inner}), solid lines] are
compared to the solutions of the mesoscopic asymptotic Eq.\,(\ref{outout}).
For the latter profiles are given that have the same $\theta_0^{adv}$
(dot-dashed lines) or the same $\delta$ (dashed line) as the numerical solution.
Parameters are (a) numerical: $\theta_0^{adv}=0.8$, $\delta=0.009$;
asymptotic (dot-dashed)  $\theta_0^{adv}=0.8$, $\delta=0.010$, $s_0=6.35$, 
$K/V^{1/2}=2.63$; asymptotic (dashed)  $\theta_0^{adv}=0.82$, $\delta=0.009$, $s_0=2.71$, 
$K/V^{1/2}=2.71$; and (b) numerical: $\theta_0^{adv}=0.6$, $\delta=0.015$;
asymptotic (dot-dashed)  $\theta_0^{adv}=0.6$, $\delta=0.019$, $s_0=3.08$, 
$K/V^{1/2}=1.98$; asymptotic (dashed)  $\theta_0^{adv}=0.69$, $\delta=0.015$, $s_0=4.16$, 
$K/V^{1/2}=2.21$.
For comparison the equilibrium profile of a droplet on a homogeneous substrate
without wettability gradient is also shown (dotted lines). 
\label{compprof}\no{compprof}
}
\end{figure}
}

The numerical results obtained by continuation (solid lines) strongly differ
from the static droplet shapes (dotted lines) that are the basis for
the  small-$\delta$ approximation Eq.~(\ref{pp04delta}). The asymptotic 
mesoscopic profiles [Eqs\,(\ref{xaia}) and (\ref{hia}) with the parameters
obtained from Eqs.\,(\ref{hph2}), (\ref{eqvol}) and (\ref{advmatch})]
approach the numerical results reasonably good for weak driving
$\theta_{adv}^0=0.8$, independently of whether one compares profiles for identical
velocity or driving (Fig.\ref{compprof}\,(a)). For larger driving, 
the comparison of profiles for identical velocities gives better results.
In general, the receding part is
described quite perfectly. The advancing part differs because the matching is
based on the advancing equilibrium contact angle that is smaller than the
dynamical one.

Our treatment has made it clear that the characteristics of the moving
droplets depend in a crucial way on the kind of driving used. The droplet may be
driven by body forces, as for instance, gravitation or Marangoni forces. In
lubrication theory the latter also takes the form of a body force althought
physically it acts at the free surface only. The driving is top-down because
the force is fed into the system on the macroscopic scale and causes motion on
all scales down to the microscale. One of our main results 
is that this type of driving cannot be described by the present theory
because the balance of the viscous term and the capillary term in 
Eq.\,(\ref{outout}) does not account for the driving force.
Specifically, it is not possible to use the solution of Eq.\,(\ref{outout})
in terms of Airy functions to describe droplets sliding down an incline driven
by gravity. This is already obvious from the fact that for gravity-driven
drops the advancing dynamic contact angle is larger than the receding one 
\cite{Thie01,Thie02} contrary to the characteristics of the mesoscopic
solution given by Eqs.\,(\ref{xaia}) and (\ref{hia}).

On the contrary, driving the droplets by a wettability gradient is bottom-up
because the force is fed into the system on the microscale and causes motion
up to the mesoscale (in our terms, no macroscale exists in this
case because the macroscale is defined by the scale of the body forces that
are absent by definition of the problem). 

\acknowledgements{
LMP and UT acknowledge support by the Israel Science Foundation (grant \#
~55/02) and the European Union (MRTN-CT-2004--005728), respectively.
}

\end{document}